\documentclass[10pt]{article}

\usepackage{epsfig}

\begin{document}

\title{\bf Study of the $\rho$, $\omega$, $\phi\to\eta\gamma\to 7\gamma$ Decays
with an SND Detector on a VEPP-2M Collider}
\author{
M.N.Achasov,
S.E.Baru,
K.I.Beloborodov,
A.V.Berdyugin\footnote{email:berdugin@inp.nsk.su}~,
A.V.Bozhenok,\\
A.G.Bogdanchikov,
D.A.Bukin,
S.V.Burdin,
A.V.Vasiljev,
Yu.S.Velikzhanin,\\
T.V.Dimova,
V.P.Druzhinin,
M.S.Dubrovin,
D.I.Ganyushin,
I.A.Gaponenko,\\
V.B.Golubev,
V.N.Ivanchenko,
I.A.Koop,
A.A.Korol,
S.V.Koshuba,\\
G.A.Kukartsev,
E.V.Pakhtusova,
A.A.Salnikov,
S.I.Serednyakov,
V.V.Shary,\\
Yu.M.Shatunov,
V.A.Sidorov,
and Z.K.Silagadze\\
~~~\\
{\it Budker Institute of Nuclear Physics,}\\
{\it Siberian Branch of the Russian Academy of Sciences} \\
{\it Lavrentyev 11, Novosibirsk, 630090, Russia}
}

\date{}
\maketitle

\begin{abstract}

The $e^+e^-\to\eta\gamma\to 7\gamma$ process was studied in the energy range
$2E=600\div 1060$ MeV with an SND detector on a VEPP-2M $e^+e^-$ collider.
The decay branching ratios
$B(\phi\to\eta\gamma)=(1.343\pm 0.012\pm 0.055)\cdot 10^{-2}$,
$B(\omega\to\eta\gamma)=(4.60\pm 0.72\pm 0.19)\cdot 10^{-4}$, and
$B(\rho\to\eta\gamma)=(2.69\pm 0.32\pm 0.16)\cdot 10^{-4}$
were measured.
\end{abstract}

Radiative decays of light vector mesons --- $\rho$, $\omega$, and $\phi$ ---
are important for understanding the behavior of the strong interaction at
low energies. Although many measurements were carried out for the probabilities
of the radiative decays, the achieved accuracy \cite{pdg} is insufficient
for reliable determination of the parameters of phenomenological models
[2-4].

We report the results of investigations of the $e^+e^- \to\eta\gamma$ process
followed by the $\eta\to 3\pi^0\to 6\gamma$ decay. Since the final state includes
seven photons, the background can be substantially suppressed compared
to that in other channels of $\eta$-meson decay and, there-fore,
the systematic error may be reduced.

The experiment \cite{link4} was carried out in 1998 at the VEPP-2M $e^+e^-$ collider
with the SND detector \cite{link6}. Two scans were performed in the energy range
$2E_0=984\div 1060$ MeV (PHI-98 experiment) with an integral luminosity of
$8.0~pb^{-1}$ at 16 energy points and with about $10^7$ produced $\phi$-mesons.
In addition, a scan (OME-98 experiment) over 38 points in the energy range
$2E_0=360\div 970$ MeV was carried out with an integral luminosity of
$3.5~pb^{-1}$ and with about $3\cdot 10^{6}$ produced $\rho$ and $\omega$  mesons.

The events of the process
\begin{equation}
\label{etag}
e^+e^-\to\eta\gamma~,~\eta\to 3\pi^0~,~\pi^0\to 2\gamma
\end{equation}
are characterized by the final state with seven photons, a few of which may
not be detected. The extra photons may also appear due to the splitting of
a shower in the calorimeter, the emission of the photons by the initial
particles at large angles, or the superposition of the beam background. The main
background process in the $\phi$-resonance region is the $\phi\to K_SK_L$ decay,
where $K_S$ decays into two neutral pions and $K_L$ , interacting in the
calorimeter, produces extra ``photons.'' An additional background is formed by the
$e^+e^-\to\omega\pi^0+X$ process followed by the $\omega\to\pi^0\gamma$ decay, where
$X$ are extra photons. An analysis of the experimental data has demonstrated
that the QED process $e^+e^-\to 3\gamma$, being superposed with other events, may
also result in the required event configuration.

Taking the above-listed background events into account, we selected events in two
steps. At the first step, among the events in which six or more photons and no charged
particles were detected, we selected those satisfying the following conditions imposed
on a total energy release $E_{tot}$ in the calorimeter and the total momentum
$P_{tot}$ of photons:
\[ E_{tot}/2E_0 < 1.2~,~P_{tot}/2E_0 < 0.2/c~,~E_{tot}/2E_0 - cP_{tot}/2E_0 > 0.7~.\]

For the selected events, we performed a kinematic reconstruction using the measured
angles, the energies of the photons, and energy-momentum conservation. As a result,
the energies of the photons were determined more accurately and the $\chi^2$ values
specifying the degree of certainty of a process were determined:
\begin{itemize}
\item $\chi^2$ for the assumption of the $e^+e^-\to n\gamma$ process
      with $n\geq 6$ or 7 ;
\item $\chi^2_{3\gamma}$ for the assumption of the $e^+e^-\to 2(3)\gamma + X$ process ;
\item $\chi^2_{\omega\pi^0}$ for the assumption of the $e^+e^-\to\omega\pi^0+X$ process .
\end{itemize}
Further selection was carried out with the restrictions
\[ \chi^2<30~,~\chi^2_{3\gamma}>20~,~\chi^2_{\omega\pi^0}>20~. \]

\begin{figure}[t]
\epsfig{figure=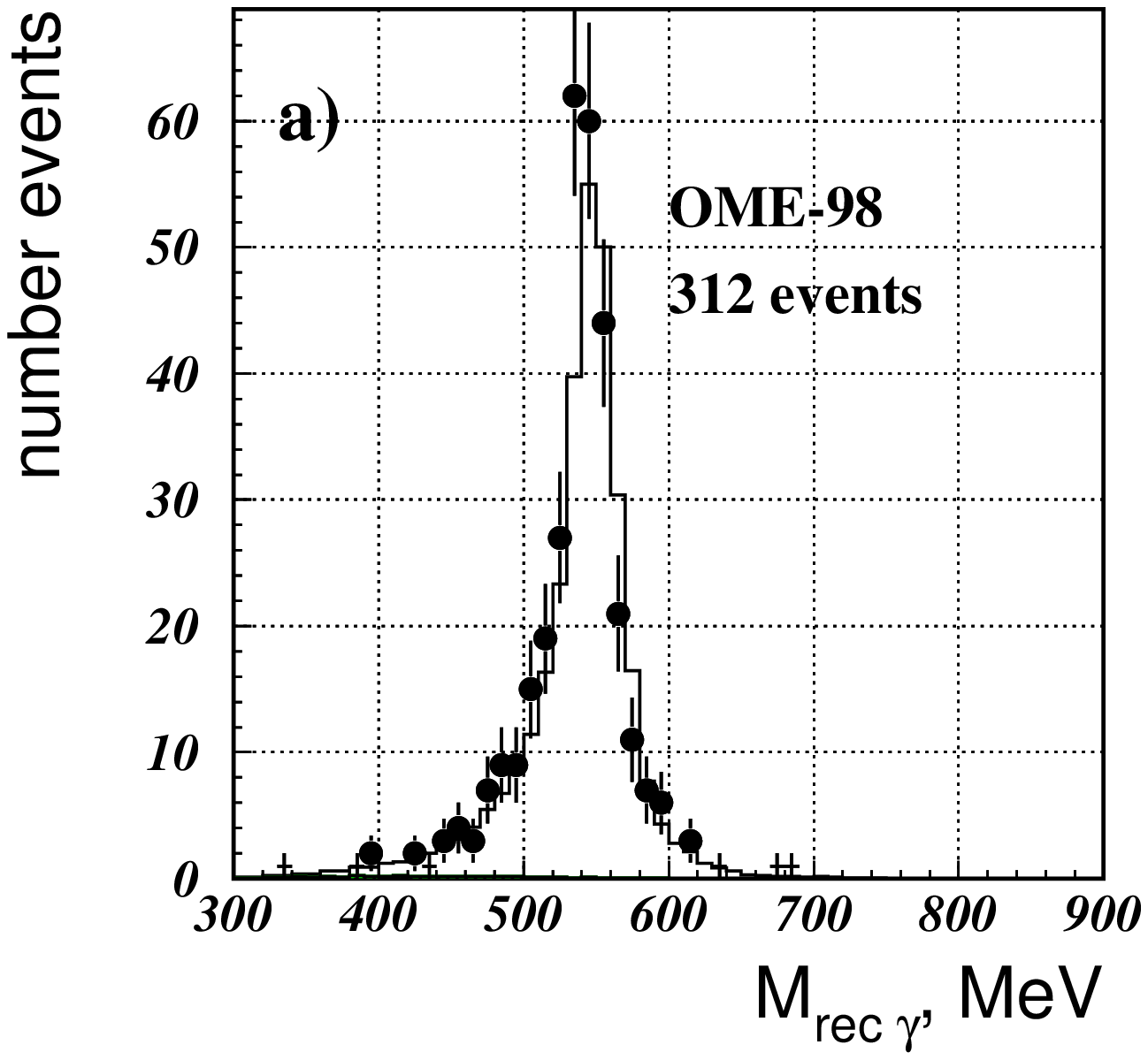,width=0.49\textwidth}
\hfill
\epsfig{figure=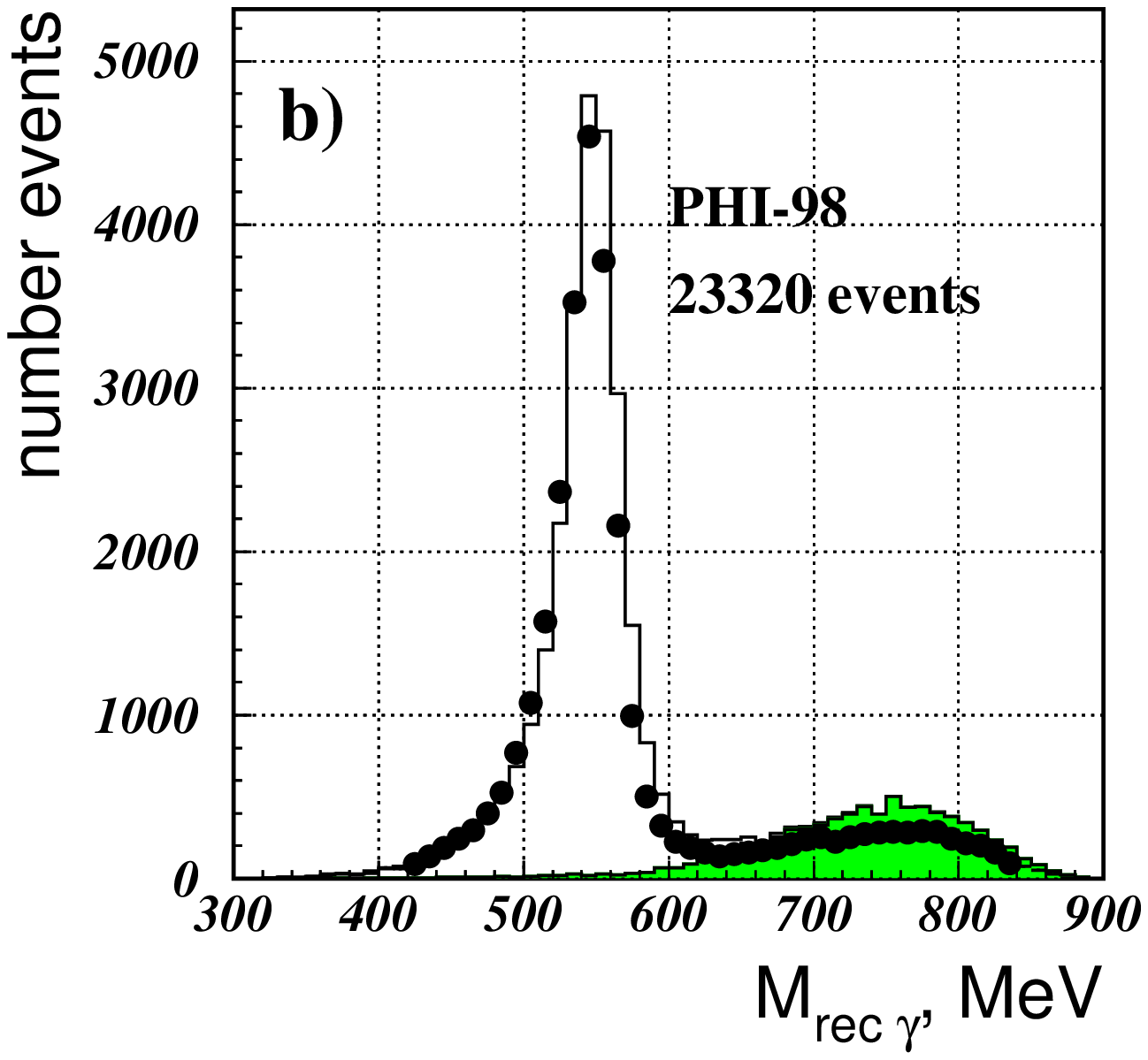,width=0.49\textwidth}
\caption{\label{ere1} Recoil mass distribution for the photon with the highest energy
in an event. The points are the experimental data, and the histograms are the
simulation: (a) OME-98 experiment (312 events) and (b) PHI-98 experiment (23320 events).
The shaded histogram is the sum of the simulation of background processes, which reduces
to process (\ref{kskl}).}
\end{figure}

Figure \ref{ere1} shows the distributions of the selected events in the recoil mass
$M_{rec~\gamma}$ of the highest energy photon. It is seen that the desired process
prevails in all scans. The events with $M_{rec~\gamma}>600$ MeV in Fig. \ref{ere1}b
are determined by the process 
\begin{equation}
\label{kskl}
e^+e^-\to\phi\to K_SK_L~.
\end{equation}
Finally, we select the events satisfying the condition $400<M_{rec~\gamma}<600$ MeV.

The number $N(s)$ of the observed events at a given energy is described by the formula
\begin{equation}
N(s) = L(s)\left[\epsilon(s)\beta(s)\sigma(s) + \sigma_b(s)\right]~,~s=4E^2_0~,
\end{equation}
where $L(s)$ is the integral luminosity, $\epsilon$ is the detection efficiency
determined by a simulation, $\beta$ is the factor representing the radiative
corrections, $\sigma_b$ is the cross section for background processes, and $\sigma$
is the cross section for the desired process (\ref{etag}).

When determining the background from process (\ref{kskl}), inaccuracy in the
simulation of the interaction of the $K_L$ meson with a substance in the calorimeter
is possible. Figure \ref{ere1}b demonstrates that the contribution of process
(\ref{kskl}) dominates for $M_{rec~\gamma}>600$ MeV, while the contribution of the
desired process (\ref{etag}) is negligible. For this reason, the number of events of
process (\ref{kskl}) in the range $400<M_{rec~\gamma}<600$ MeV was determined from the
number of experimental events in the interval $600<M_{rec~\gamma}<800$ MeV by taking
into account the simulated ratio of the numbers of the $K_SK_L$ events that fall into
the mass ranges $400<M_{rec~\gamma}<600$ MeV and $600<M_{rec~\gamma}<800$ MeV.

\begin{figure}[t]
\epsfig{figure=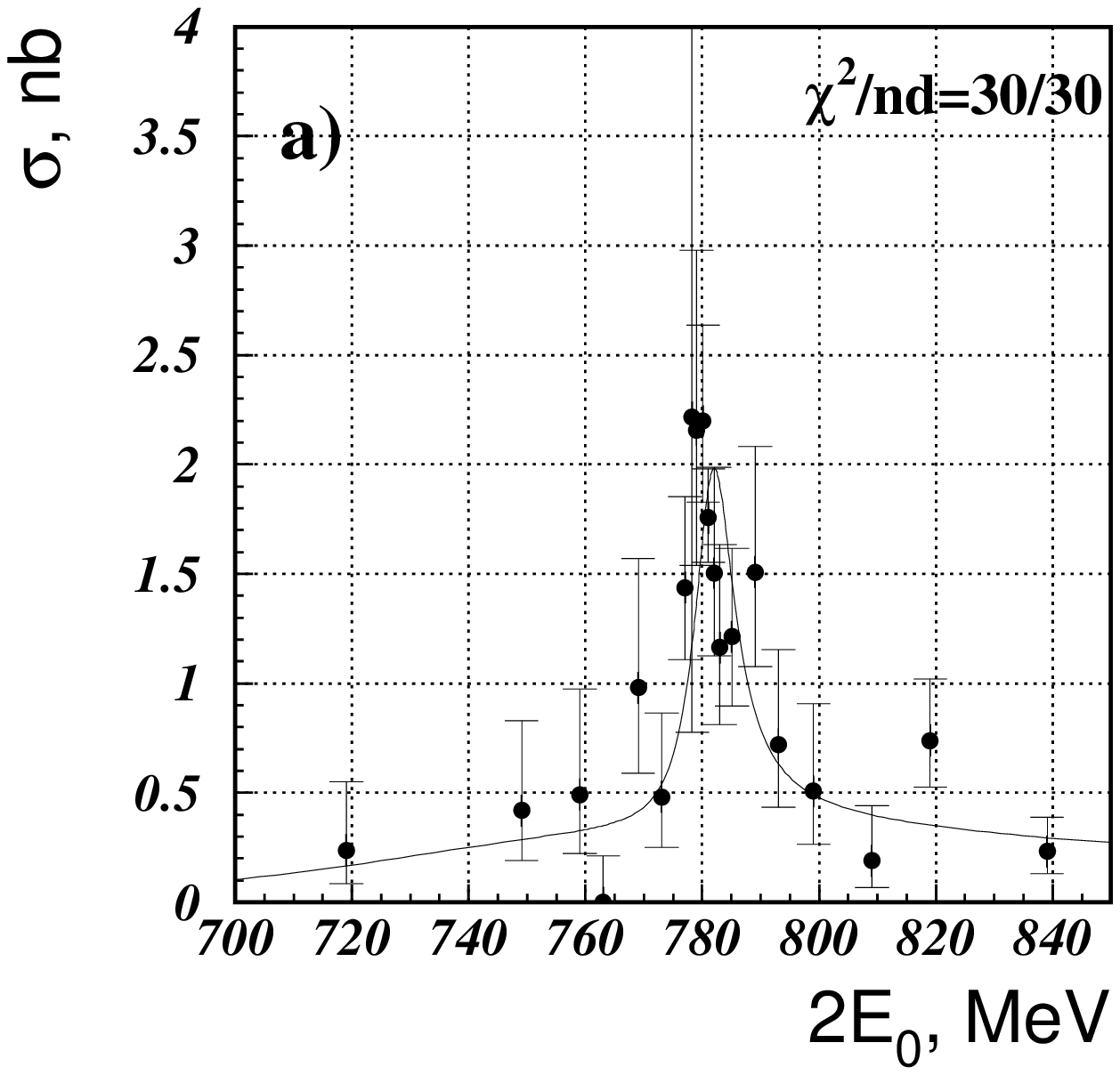,width=0.49\textwidth}
\hfill
\epsfig{figure=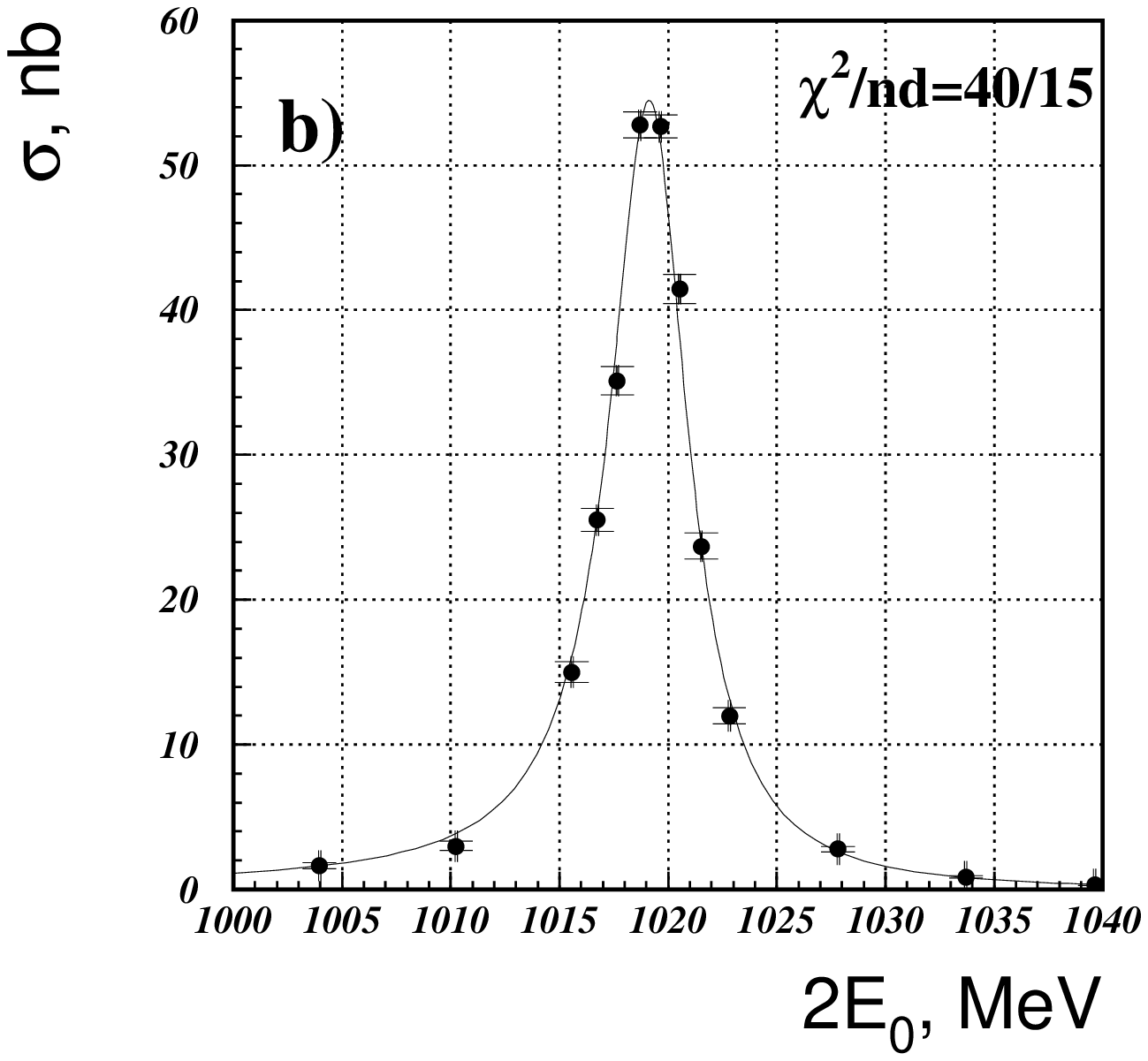,width=0.49\textwidth}
\caption{\label{crsec} Measured total cross section for the $e^+e^-\to\eta\gamma$
process in the region of (a) $\rho$ and $\omega$ mesons ($\chi^2/nd = 30/30$) and
(b) $\phi$ mesons ($\chi^2/nd = 40/15$).}
\end{figure}

The energy dependence of the resulting cross section (Fig. \ref{crsec}) was parametrized
by the vector-dominance formulas \cite{link7} including the contributions of the
$\rho$, $\omega$, and $\phi$ resonances:
\begin{equation}
\sigma(s)~=~\frac{F(s)}{s^{3/2}}\left|\sum\limits_{V=\rho,\omega,\phi}\sqrt{12\pi\prod\limits_{B_{VP}}\frac{m^3_V}{F(m^2_V)}}\frac{\Gamma_Ve^{i\varphi_V}}{D_V(s)}\right|^2~,
\end{equation}
where $F(s)=[(s-m^2_{\eta})/2\sqrt{s}]^3$, $D(s)=m^2_V-s-i\sqrt{s}\Gamma_V(s)$,
and the products
\[ \prod\limits_{B_{VP}}=Br_{V\to e^+e^-}\cdot Br_{V\to\eta\gamma}\cdot Br_{\eta\to 3\pi^0}\cdot Br^3_{\pi^0\to 2\gamma} \]
are the free parameters of the approximations. The relative phase shifts of the
resonances were fixed at $\varphi_{\rho}=\varphi_{\omega}=0$, $\varphi_{\phi}=\pi$. The
approximation gives the following results:
\begin{eqnarray}
\label{br3v}
\prod\limits_{B_{\phi P}} & = & (1.249 \pm 0.011\pm 0.035)\cdot 10^{-6}~, \nonumber \\
\prod\limits_{B_{\omega P}} & = & (1.01\pm 0.16\pm 0.03)\cdot 10^{-8}~, \nonumber \\
\prod\limits_{B_{\rho P}} & = & (3.77\pm 0.45\pm 0.11)\cdot 10^{-9}~.
\end{eqnarray}
where the first error is statistical and the second, systematic, error is due to the
contributions of errors in the determination of the detection efficiency and the error
in the measurement of the luminosity. The luminosity was measured from elastic
electron-positron scattering at large angles and from the process of two-quantum
annihilation. The difference in the results of the two methods does not exceed 1\%.
The accuracy of the theoretical formulas used for the simulation of elastic scattering
and experimental conditions provides an estimate of about 2\% for the accuracy of the
luminosity in the case under consideration. In order to estimate the systematic errors
in the detection efficiency, the stability of the results to a change in the selection
conditions was examined: we added restrictions on the polar angle of the photons and on
the number ($N_{\gamma}=7$) of particles and used only the completely reconstructed
$e^+e^-\to\eta\gamma\to 7\gamma$ events. As was discussed above, due to the emission
of the initial particles at large angles and superposition of preceding events, extra
spurious photons appear in the SND calorimeter. For this reason, one of the tests of
the kinematic reconstruction was carried out with the exclusion of photons with energies
lower than 50 MeV and with a polar angle less than $36^\circ$. In addition, two scans
of the $\phi$ meson were independently processed. All tests demonstrated the stability
of the results, and the total systematic error in the efficiency, with the inclusion of
all effects, was estimated at 2\%. This estimate is treated as independent of the
systematic error in luminosity.

Using the tabular values of $Br_{V\to e^+e^-}$, $Br_{\eta\to 3\pi^0}$,
and $Br_{\pi^0\to 2\gamma}$ from \cite{pdg}, we obtain from Eqs. (\ref{br3v})
the values
\begin{eqnarray}
\label{brsv}
Br_{\phi\to~e^+e^-}~\cdot~Br_{\phi\to\eta\gamma} & = & (4.017\pm 0.035\pm 0.124)\cdot 10^{-6}~, \nonumber \\
Br_{\omega\to~e^+e^-}~\cdot~Br_{\omega\to\eta\gamma} & = & (3.25\pm 0.51\pm 0.10)\cdot 10^{-8}~, \nonumber \\
Br_{\rho\to~e^+e^-}~\cdot~Br_{\rho\to\eta\gamma} & = & (1.21\pm 0.14\pm 0.04)\cdot 10^{-8}~, \nonumber \\
Br_{\phi\to\eta\gamma} & = & (1.343\pm 0.012\pm 0.055)\cdot 10^{-2}~, \nonumber \\
Br_{\omega\to\eta\gamma} & = & (4.60\pm 0.72\pm 0.19)\cdot 10^{-4}~, \nonumber \\
Br_{\rho\to\eta\gamma} & = & (2.69\pm 0.32\pm 0.16)\cdot 10^{-4}~, \nonumber \\
\sigma_{\phi\to\eta\gamma} & = & (56.75\pm 0.50\pm 1.68)~nb~, \nonumber \\
\sigma_{\omega\to\eta\gamma} & = & (0.78\pm 0.12\pm 0.02)~nb~, \nonumber \\
\sigma_{\rho\to\eta\gamma} & = & (0.300\pm 0.036\pm 0.009)~nb~,
\end{eqnarray}
where $\sigma_{V\eta\gamma}=12\pi Br_{V\to e^+e^-}Br_{V\to\eta\gamma}/m^2_V$
and the errors of the tabular values are included in the systematic errors. Using
data (\ref{br3v}) obtained above, the results of previous measurements with the SND
detector \cite{link10,link8}, and  the data on the widths of the
$\rho$ and $\omega$ mesons \cite{pdg}, we derive the ratios
\begin{eqnarray}
\label{otn}
\frac{B(\eta\to\pi^+\pi^-\pi^0)}{B(\eta\to\gamma\gamma)+B(\eta\to 3\pi^0)} &=& 0.304\pm 0.012~, \nonumber \\
\frac{B(\eta\to 3\pi^0)}{B(\eta\to\gamma\gamma)} &=& 0.826\pm 0.024~, \nonumber \\
\frac{\Gamma_{\rho\to\eta\gamma}}{\Gamma_{\omega\to\eta\gamma}} &=& 10.5\pm 2.1~.
\end{eqnarray}

Our results
(5)-(7)
are in agreement with the data of
other experiments
[8-12].
The branching ratios for
the $\phi$, $\omega\to\eta\gamma$ decays are measured with an accuracy close to the
tabular one \cite{pdg}, and the branching ratio for the $\rho\to\eta\gamma$ decay is
determined with the doubly improved accuracy. Note that the quantity
$Br_{\phi\to e^+e^-}\cdot Br_{\phi\to\eta\gamma}\cdot Br_{\eta\to 3\pi^0}$ (\ref{br3v})
was measured with noticeably higher accuracy than the branching ratio
$Br_{\phi\to\eta\gamma}$ (\ref{brsv}), because the leptonic width of the $\phi$ meson
is known with an accuracy of 2.7\%, which is noticeably worse than the statistical
accuracy of our measurement.

This work was supported by the ``Russian Universities'' Foundation
(project no. 3N-339-00) and the Russian Foundation for Basic Research
(project nos. 99-02-16813, 00-02-17478, and 00-02-17481).

\end{document}